%% file: main.tex
\documentclass[%
 reprint,
superscriptaddress,
 aps,
 pra
]{revtex4-2}

\usepackage{graphicx}
\usepackage{dcolumn}
\usepackage{bm}
\usepackage[sort&compress]{natbib}
\usepackage{amsmath,bm}
\usepackage{xspace}
\usepackage{amssymb}
\usepackage{hyperref}
\usepackage{siunitx}
\usepackage[a4paper,margin=2cm]{geometry}
\usepackage{multirow}
\usepackage{physics}
\usepackage{accents}

\newcommand{\Supp}[1]{\cite{SupplementalMaterials}} 
\newcommand{\SectionPRL}[1]{\emph{#1.---~} }

\begin{document}

\renewcommand{\bibnumfmt}[1]{[#1]}
\renewcommand{\citenumfont}[1]{#1}

\title{Quantum Vacuum Nonlinearities in Laser Interferometers}

\author{Zain Mehdi}
 \email{zain.mehdi@anu.edu.au}
 \affiliation{Department of Quantum Science and Technology and Department of Fundamental and Theoretical Physics, Research School of Physics, Australian National University, Canberra 2600, Australia}%
  \author{Joseph J. Hope}
 \affiliation{Department of Quantum Science and Technology and Department of Fundamental and Theoretical Physics, Research School of Physics, Australian National University, Canberra 2600, Australia}%
 \author{Simon A. Haine}
 \affiliation{Department of Quantum Science and Technology and Department of Fundamental and Theoretical Physics, Research School of Physics, Australian National University, Canberra 2600, Australia}%

\date{\today}

\begin{abstract}
We propose all-optical tests of photon-photon interactions in the matter vacuum using standing-wave laser interferometers, which do not require external magnetic fields and use conventional laser sources. We show the interferometric detection of photon-photon scattering predicted by quantum electrodynamics is within reach of laboratory-scale experiments with a sensitivity that improves nonlinearly with the circulating power within the cavity. We outline how such an experiment could be adapted to probe properties of quantum fields beyond the Standard Model of particle physics.
\end{abstract}


\maketitle
\begin{figure}
    \centering
    \includegraphics[width=\linewidth]{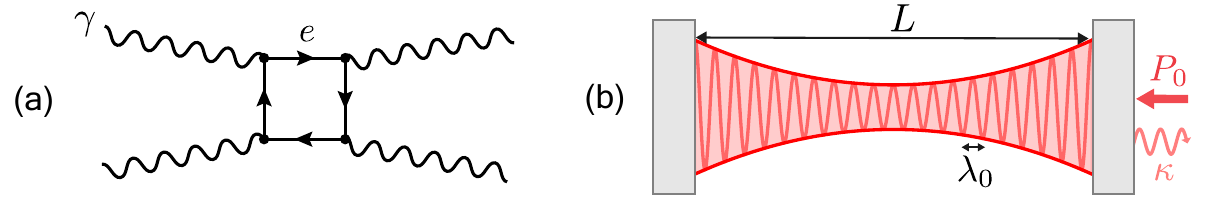}
    \caption{(a) Diagram for photon-photon interactions mediated by virtual electron-positron pairs in QED. (b) Fabry-Perot cavity formed by two mirrors separated by a length $L$, driven by input laser power $P_0$, where the optical field forms a standing wave with wavelength $\lambda_0$. Signatures of photon-photon interactions are encoded in the phase of the field transmitted from the cavity at rate $\kappa$. } 
   \label{fig:Schematic}
\end{figure}

Quantum electrodynamics (QED) predicts effective interactions between photons mediated by zero-point fluctuations of the vacuum (Fig.~\ref{fig:Schematic}a), manifesting in nonlinear corrections to Maxwellian electrodynamics at low energies~\cite{Heisenberg1936,Klein1964,Karplus1951}. Although photon-photon interactions have been observed in strong nuclear fields produced by heavy ion collisions~\cite{ATLASCollaboration2017}, they have yet to be detected in the matter vacuum. An experiment probing low-energy photon-photon scattering would enable tests of fundamental physics~\cite{Fedotov2023}, including Lorentz invariance~\cite{Marklund2009}, and could constrain properties of theorized `beyond-Standard-Model' particles such as axions~\cite{Villalba-Chavez2013}, millicharged fermions~\cite{Villalba-Chavez2016}, and gravitons~\cite{Mehdi2023_QG}, thereby complementing constraints from high-energy particle-collider experiments~\cite{Jaeckel2010}.

To date, experimental efforts have focused on vacuum birefringence due to the scattering of photons from strong magnetic fields~\cite{Cadene2014,Ejlli2020}, and are limited by the strength and stability of the external field. In particular, the PVLAS collaboration has placed experimental constraints on magnetic vacuum birefringence to within an order of magnitude of the QED prediction~\cite{Ejlli2020}. Future magnetic vacuum birefringence experiments plan to use the superconducting magnets at CERN~\cite{Ahmadiniaz2025}. Other proposals involve scattering of a laser beam from one or several ultra-intense pump fields~\cite{Gies2009,Turcu2016}, using either petawatt pulses of near-infrared light~\cite{Turcu2019,Fedeli2021} or $\mathcal{O}(100)$TW pulses from x-ray free-electron lasers~\cite{Dunne2009}, achievable in specialized facilities. 

This Letter proposes an alternative interferometric approach for resolving photon-photon interactions based on conventional laser sources without the need for external fields. Specifically, we show theoretically that interferometric detection of photon-photon interactions intrinsically present in standing-wave optical cavities is within reach of current experimental capabilities. When the cavity is held at its minimum volume, the sensitivity of the interferometric signal is independent of cavity length and scales nonlinearly with optical power. We outline differential measurement schemes to isolate optical vacuum birefringence from technical noises, and further explore the microscopic structure of the interaction, which allows the QED effect to be discriminated from photon-photon interactions mediated by virtual particles beyond the Standard Model of particle physics.

\SectionPRL{Photon-photon interactions} A generic Hamiltonian or Lagrangian density describing photon-photon interactions can be constructed from the the two Lorentz invariants of the electromagnetic (EM) field tensor, $\mathcal{G}_1=(E_iE_i-c^2B_iB_i)/2$ and $\mathcal{G}_2=cE_iB_i$ ($i=1,2,3$), where $E_i$ ($B_i$) is the $i$-th component of the electric (magnetic) field and repeated indices are implicitly summed. For the specific case of photon-photon interactions in QED, the interaction Hamiltonian density can be obtained the leading-order contribution of the Euler-Heisenberg effective action in the low energy limit~\cite{SupplementalMaterials}:
\begin{align}
\label{eq:HamiltonianDensity}
	\mathcal{H}_{\rm int} &= -\frac{2\alpha^2\varepsilon_0^2\hbar^3}{45m_e^4c^5}(4 \mathcal{G}_1^2 + 7\mathcal{G}_2^2) \,,
\end{align}
where $m_e$ is the electron mass, $\varepsilon_0$ is the permittivity of free space, $c$ is the speed of light, and spatiotemporal arguments have been suppressed. While this interaction vanishes for monochromatic plane waves, it is non-zero for standing-waves where $E_i \propto \sin(k z)$ as is realized by the optical field within a Fabry-Perot cavity of length $L$ (see Fig.~\ref{fig:Schematic}b), which has resonances for $k_n =n\pi/L$ ($n=1,2,\dots$). We will assume this geometry for the remainder of this work.

To obtain the quantized interaction Hamiltonian within a perturbative framework, it is sufficient to substitute the quantized EM field in the free-field limit into $\mathcal{H}_{\rm int}$, and integrate over spatial dimensions~\cite{SupplementalMaterials}. This calculation is simplified considerably for the EM field within an optical cavity, where we need only consider leading-order contributions from a small number of modes with macroscopic occupation, within a rotating-wave approximation. Further details are provided in Appendix \ref{app:HamiltonianDerivation}. In this work we will focus on resonant photon-photon interactions with at most two macroscopically occupied cavity modes. 

\SectionPRL{Single-mode forward scattering} We first consider the case of a single-mode standing wave with resonant frequency $\omega_0$ and polarization vector $\mathbf{e}=\{e_x,e_y,0\}$. Within a rotating-wave approximation, the leading order photon-photon interaction is the forward-scattering process described by a Kerr-like interaction Hamiltonian:
\begin{align}
\label{eq:Hint_SM}
	\hat{H}_{\rm SM} = -\left(1+\frac{3}{8}\mathcal{V}^2\right)\frac{32\alpha^2 }{45 }\frac{\lambdabar_e^3}{V}\frac{(\hbar \omega_0)^2}{m_e c^2} \hat{a}^\dag \hat{a}^\dag \hat{a} \hat {a}
\end{align}
where $\mathcal{V}=-2\Im{e_x^* e_y}$ is the circular Stokes parameter characterizing the helicity of the optical mode (see Appendix \ref{app:HamiltonianDerivation} for a detailed derivation of this equation). Note that the contribution of the two Lorentz variants to this expression are distinct, as $\mathcal{G}_2\propto \mathcal{V}$ vanishes in the case of linear polarization (see Appendix \ref{app:HamiltonianDerivation}). 

For large mode occupation, $\bar{n}\gg 1$, one can perform a mean-field analysis of $\hat{H}_{\rm SM}$ by expanding $\hat{n}= \bar{n}+\delta\hat{n}$ and retaining only the leading order term in this expansion. This leads to $\hat{a}^\dag\hat{a}^\dag \hat{a}\hat{a} \approx 2\bar{n}\hat{a}^\dag \hat{a}$, such that $\hat{H}_{\rm SM}$ describes a decrease in the eigenenergy proportional to  $\bar{n}\left(1+3\mathcal{V}^2/8\right)(\hbar \omega_0)^2/V$. This is equivalent to a nonlinear shift in the vacuum index of refraction:
\begin{align}
\label{eq:}
	n_{\rm ref} \approx 1 +\bar{n}\left(1+\frac{3}{8}\mathcal{V}^2\right) \frac{64\alpha^2}{45}\frac{\lambdabar^3}{V}\frac{\hbar \omega_0}{m_e c^2}\,.
\end{align}
This results in a measurable phase shift on the light transmitted from the cavity of approximately $\phi = (n_{\rm ref}-1)\omega_0 \tau$, where $\tau$ is the effective storage time of the cavity. For a cavity of length $L$ and finesse $\mathcal{F}$, $\tau$ may be estimated as the product of a single round-trip time $2L/c$ and the total number of round trips $2\mathcal{F}/\pi$. Assuming the cavity to be overcoupled with bandwidth $\kappa=\pi c/(\mathcal{F} L)$~\cite{Aspelmeyer2014b}, this is equivalent to $\tau = 4/\kappa$. Using $\bar{n}\approx4P_0/(\hbar\omega_0\kappa)$ to express the phase shift in terms of the input laser power $P_0$~\cite{Aspelmeyer2014b},	
\begin{align}
\label{eq:PhaseShift}
\phi&\approx 	 \left(1+\frac{3}{8}\mathcal{V}^2\right)\frac{1024\alpha^2}{45 \kappa^2}\frac{\lambdabar^3}{V}\frac{P_0\omega_0}{m_e c^2} \,.
	\end{align}
To maximize this phase shift, we assume the cavity volume can be held near its minimum volume $V$, which is fundamentally limited by laser beam divergence to approximately $V_{\rm min}\approx \lambda_0 L^2/\sqrt{3}$~ corresponding to a beam waist of $w_0 \propto \sqrt{\lambda_0 L}$ for a Gaussian beam (see Appendix B). 

Notably, the phase shift is \emph{independent} of the cavity length $L$ for $V= V_{\rm min}$ as $\kappa\propto L^{-1}$, which is distinct from existing interferometers looking for vacuum birefringence induced by external magnetic fields~\cite{Cadene2014,Ejlli2020,Ballou2018}. In addition to allowing laboratory-scale experiments, the length-independence of the QED phase shift means it can be discriminated from effects due to high optical intensity at the mirror surface (e.g. thermo-elastic distortion) which scale inversely with $L$ for the above choice of beam waist. 

Assuming the phase to be measured continuously for time $T\gg \tau$, photon shot noise sets a limit to the achievable precision (without squeezing or non-Gaussian protocols, which we consider later) of $\Delta \phi_{\rm SN} =\sqrt{\tau/(\bar{n}T)}\approx \sqrt{\hbar \omega_0/ (P_0 T)}$. Therefore, by averaging the monitored phase of the transmitted light field from the cavity over a total time $T$ at the shot-noise limit, one could achieve a signal-to-noise ratio (SNR) that is independent of $L$ and scales non-linearly with the cavity finesse and input power:
	\begin{align}
	\label{eq:PhaseShift_SNR}
		\frac{\phi}{\Delta \phi_{\rm SN}} &\approx \frac{1024\left(1+\frac{3}{8}\mathcal{V}^2\right)\alpha^2}{15\pi^3\sqrt{3\hbar}}\frac{\lambdabar^3}{m_e c^5}\mathcal{F}^2(\omega_0 P_0)^{3/2}\sqrt{T} \,, \\
		&\approx \left(\frac{\mathcal{F}}{{7\times 10^5}}  \right)^{2}\left(\frac{P_{0}}{1\,{\rm W}}\frac{1\mu {\rm m}}{\lambda_0}\right)^{3/2}\left( \frac{T}{1\,{\rm day}} \right)^{1/2} \,. \notag
	\end{align}
	In the second line we have assumed a linear polarization such that $\mathcal{V}=0$, and chosen parameters compatible with experimentally-demonstrated cavity finesse~\cite{Ejlli2020} and circulating power $P_{\rm circ}\approx \mathcal{F}P_0\approx 700$kW~\cite{Lu2024}. This demonstrates the QED nonlinearity could feasibly be resolved with continuous shot-noise-limited measurements over a $24$-hour period in a laboratory-scale experiment, without the need for external fields.

 Eq.~\eqref{eq:PhaseShift_SNR} is a standard quantum limit for the detection of QED forward scattering. In principle, this sensitivity could be surpassed by beyond-mean-field protocols that can achieve Heisenberg scaling ${\rm SNR} \sim P_0^2$~\cite{Mehdi2023_QG}. We will not discuss this further here as we will focus primarily on interferometric readouts feasible with current technologies. A caveat of this calculation is the assumption that the phase noise is dominated by shot noise, rather than radiation pressure effects which typically dominate high-power interferometers at low frequencies~\cite{Danilishin2012}. This assumption could be realized by modulating the power of the input laser at frequency $\Omega$ to shift the phase signal from DC to a tunable AC band where radiation pressure effects are suppressed below shot noise by the mechanical susceptibility of the cavity mirrors, e.g. above roughly $100$Hz for mirror masses of $\mathcal{O}(10)$kg. This requires the bandwidth of the cavity to be greater than several hundred hertz, which is feasible for cavity lengths of $\mathcal{O}(1)$m when the cavity mode is at its minimum volume. The freedom to choose the  modulation frequency also allows the QED effect to be distinguished from various technical noises through their distinct frequency responses.

The polarization dependence of Eq.~\eqref{eq:Hint_SM} allows one to eliminate certain background signals. For example, one could compare the interferometric signal between two cavities (e.g. in a Michelson arrangement) where one hosts a linear polarized beam ($\mathcal{V}=0$) and the other hosts a circular polarized beam ($\mathcal{V}^2=1$). Assuming equal power in each cavity, the two cavities should accumulate a differential phase that is smaller than the single-cavity shift by a factor of $8/3\approx 2.7$. This should vanish if both beams are linearly polarized. However, it may be more convenient to exploit the richer structure of interactions between two macroscopically occupied modes of the same cavity, where it should be possible to reject spurious signals that are common to both components such as path length variations.

\SectionPRL{Cross-phase modulation and birefringence} Specifically, we consider interactions between two modes with wavevectors $\omega_1 = n_1\pi c /L$ and $\omega_2 = n_2\pi c/L$ with complex polarization vectors $\mathbf{e}_1$ and $\mathbf{e}_2$. In general the two modes may be non-degenerate with non-orthogonal polarizations. We focus on the resonant two-mode interaction which describes cross-phase modulation, for which we derive the following interaction Hamiltonian of the form $\hat{H}_{\rm TM}=-\hbar \lambda(\mathbf{e}_{1},\mathbf{e}_{2}) \hat{n}_1\hat{n}_2$, (see Appendix \ref{app:HamiltonianDerivation}), where $\hat{n}_j$ is the photon number operator for the $j$-th mode ($j=1,2$) and $\lambda$ is the per-particle interaction rate:
\begin{align}
	\lambda(\mathbf{e}_{1},\mathbf{e}_{2})=  \mathcal{A}_{\rm pol}\frac{8\alpha^2}{45}\frac{ \lambdabar^3}{V}\frac{\hbar\omega_1\omega_2}{m_e c^2} \,.
\end{align}
Here $\mathcal{A}_{\rm pol}$ encodes the dependence of the interaction on the overlaps $a=\left| \mathbf{e}_1^*\cdot\mathbf{e}_2\right|^2$ and $b=| \mathbf{e}_1\cdot\mathbf{e}_2| ^2 $ between the polarization vectors for each mode $\mathbf{e}_{1,2}$ (each normalized as $\left| \mathbf{e}_j^*\cdot\mathbf{e}_j\right|^2=1$), 
\begin{align}
\label{eq:TM_PolarizationDependence}
  \mathcal{A}_{\rm pol}&=14-3a-3b + \delta_{n_1,n_2}\left(14a-3(b+1) \right)
\end{align}
Clearly, the two-mode interaction Eq.~\eqref{eq:Hint_SM} can distinguish linear and elliptical polarizations, as in the latter case $a\neq b$. Furthermore, $\mathcal{A}_{\rm pol}$ is independent of the frame of reference used to define the two polarization vectors and does not distinguish between which beam carries which polarization in the degenerate case. Mapping out these null dependences should allow the two-mode interaction to be isolated from technical effects such as mirror birefringence, which typically have a preferred polarization axis.

The factor of $\delta_{n_1,n_2}$ in Eq.~\eqref{eq:TM_PolarizationDependence} encodes whether the two modes are degenerate ($\delta_{n_1,n_2}=1$) or non-degenerate ($\delta_{n_1,n_2}=0$). Examples of the former case include two orthogonal polarizations of a monochromatic field (in which case $a=0$). Nearly-degenerate modes with non-orthogonal polarizations could be realized by a bichromatic field resonant with two neighboring resonances of the cavity, for example. One could also consider the case of two overlapping optical fields of two distinct cavities with the same frequency, in which case $\delta_{n_1,n_2}=1$ and $ \mathcal{A}_{\rm pol}$ should be rescaled by the normalized intensity overlap between the two cavity fields, i.e. $ \mathcal{A}_{\rm pol}\rightarrow \eta \mathcal{A}_{\rm pol}$ where $\eta \approx \int d^3 r I_1(r)I_2(r)/[(\int d^3 r I_1(r))(\int d^3 r' I_2(r'))]$ and $I_j(r)$ is the intensity profile of the $j$-th cavity field. 

The two-mode interaction Hamiltonian $\hat{H}_{\rm TM}=-\hbar \lambda(\mathbf{e}_{1},\mathbf{e}_{2}) \hat{n}_1\hat{n}_2$ leads to phase shifts of each mode proportional to the number of photons in the other, i.e. cross-phase modulation. For simplicity, the following analysis will be restricted to the limiting case of two non-degenerate modes of the same cavity with (non-orthogonal) linear polarizations, such that $\lambda(\mathbf{e}_{1},\mathbf{e}_{2}) \rightarrow \lambda(\Theta)\propto 14-6\cos^2\Theta$ where $\Theta$ is the angle enclosed by the two vectors $\mathbf{e}_1$ and $\mathbf{e}_2$. In this case, the parallel [$\Theta=0$] and perpendicular [$\Theta = \pi/2$]  polarization components of each mode (with respect to the other) will accumulate a differential phase over the optical storage time of the cavity proportional to:
\begin{align}
	\delta &= \lambda(\pi/2)-\lambda(0)= \frac{48\alpha^2}{45}\frac{ \lambdabar^3}{V_{\rm min}}\frac{\hbar \omega_1\omega_2}{m_e c^2} \,.
\end{align}
In other words, each frequency component will develop a small ellipticity due to \emph{optical} vacuum birefringence. The differential phase shift could be directly read out by choosing the two polarization vectors to have a relative angle of $\pi/4$ (i.e. $\mathbf{e}_{1}\cdot \mathbf{e}_{2}=1/\sqrt{2}$) and monitoring the relative phase between polarization components of each mode. For instance, consider the case where one mode is diagonally polarized and the other horizontally polarized, i.e. $\mathbf{e}_1=(1,1,0)/\sqrt{2}$ and $\mathbf{e}_2=(1,0,0)$. From the perspective of the first mode, the second mode acts as a horizontally-polarized pump, and so the horizontal and vertical polarization components of the first mode will accrue a differential phase proportional to $\bar{n}_2\delta$ (where $\bar{n}_j\equiv \langle \hat{n}_j\rangle$). The differential phase could be read out using homodyne detection, e.g. by choosing the polarization of the local oscillator to be orthogonal to the polarization of the carrier such that it only beats with the QED-induced phase sidebands. The same reasoning applies for the second mode which sees the first mode as a diagonally-polarized pump, such that the diagonal and anti-diagonal polarization components of the second mode will also accrue a relative phase shift proportional to $\bar{n}_1\delta$.

To analyse the sensitivity of the proposed experiment to the differential signal, we calculate the noise power spectral density (PSD) associated with estimating the differential per-particle interaction rate $\delta$ using input-output theory of optical cavities (see Appendix \ref{app:NoisePSD}). Assuming shot-noise-limited detection and a squeezed light source, the noise in the estimated value of $\delta$ from the phase signal on the $i$-th frequency component induced by the occupation of mode $j$ is described by the PSD:
\begin{align}
\label{eq:PSD_ij}
    S_{\delta\delta}^{ij} \approx \kappa\left(1+\frac{\Omega_j^2}{(\kappa/2)^2}\right)\frac{e^{-2r}}{16\bar{n}_{i} \bar{n}_{j}^2 } \,,
\end{align}
where we have neglected the contribution of optical losses at the cavity walls, which will restrict the achievable level of phase squeezing (set by the squeezing parameter, $r>0$). We assume both modes are monitored with equal populations, and the estimated value of $\delta$ from the phase readout of each of the two frequency components is averaged. In this case, the combined noise PSD is $S_{\delta\delta} = \sqrt{|S_{\delta\delta}^{12}|^2 +|S_{\delta\delta}^{21}|^2}/4$. If only one mode was measured, Eq.~\eqref{eq:PSD_ij} implies that it would be optimal to have one third of the cavity photons in the measured mode (and the remaining in the `pump' mode) for a fixed total circulating power in the cavity. Provided the $\Omega_j \ll \kappa/2$, the total phase noise integrated over a time $T$ scales as $(\Delta \delta)^2\equiv {\rm Var}(\delta)= S_{\delta\delta}/T \propto e^{-2r}P_{\rm circ}^{-3}(\mathcal{F} T)^{-1}$, which gives the same SNR scaling as the single-mode case given by Eq.~\eqref{eq:PhaseShift_SNR}. 

For the same experimental parameters used in Eq.~\eqref{eq:PhaseShift_SNR}: $P_0=1$W of $\approx 1\mu$m light (assuming $\bar{n}_1=\bar{n}_2$) and $\mathcal{F}=7\times 10^5$, the expected quantum noise at frequencies large enough to suppress radiation pressure noise (but less than the cavity half-bandwidth, $\kappa/2$) is $\sqrt{S_{\delta\delta}} \approx 10^{-23} \sqrt{\rm Hz}$ without squeezing or $6\times 10^{-24} \sqrt{\rm Hz}$ with $7$dB of squeezing ($r\approx 0.81$). For the same parameters, the birefringent parameter is approximately $\delta \approx 2\times 10^{-26}$Hz. Thus, to resolve optical vacuum birefringence, the proposed experiment requires an interrogation time of at least four days without squeezing or one day with $7$dB of phase squeezing in order to achieve $\delta\gtrsim\Delta\delta$. For a slightly larger intra-cavity power of one megawatt ($P_0\approx 1.4$W for the above finesse), the interrogation time could be reduced by a factor of approximately $(1.4)^3\approx 2.9$ as the sensitivity to the nonlinear phase shift scales as $1/\sqrt{P^3 T}$.

In principle, the sensitivity given by Eq.~\eqref{eq:PSD_ij} can be surpassed using nonlinear readouts with highly correlated optical states. To see this, we consider the quantum Fisher information $\mathcal{I}_Q$ associated with the encoding of $\lambda$ on a two-mode squeezed-vacuum state $\ket{\rm TMSV}=\sum_n (-e^{i\phi}\tanh(r))^n/\cosh(r) |n,n\rangle$, where $r$ and $\phi$ are the squeezing amplitude and phase. In Appendix \ref{app:QFI} we derive the expression $\mathcal{I}_Q=\sinh ^2(r) (\cosh (2 r)+3 \cosh (6 r))\approx 80N^4$ where $N= \langle \hat{n}_1+\hat{n}_2\rangle =  \sinh^2(r) \gg 1$ is the total photon number in the cavity. Appealing to the quantum Cramer-Rao bound, we can place an approximate upper bound on the achievable signal-to-noise ratio on measuring $\lambda$ that scales quadratically with the mean photon number, i.e. ${\rm max(SNR)}\propto P_{\rm circ}^2 \sqrt{\omega_1\omega_2\mathcal{F} L T}$ where $P_{\rm circ}$ is the total circulating power in the cavity. We include this result for completeness as it illustrates the fundamental precision limit of such an experiment is well beyond what is considered above; however we note that achieving Heisenberg scaling for large photon fluxes is currently beyond experimental capabilities.

\SectionPRL{Mediators beyond  the Standard Model} While we have focused on nonlinearities of QED, photon-photon interactions could be mediated by virtual contributions of quantum fields beyond the Standard Model of quantum physics, e.g. by millicharged fermions~\cite{Villalba-Chavez2016}, axion-like particles~\cite{Villalba-Chavez2013,Evans2019}, or even gravitons~\cite{Mehdi2023_QG}. The proposed experiments could therefore probe the properties of these postulated fields purely through their zero-point fluctuations (i.e. even in the absence of `real' particles in the initial and final conditions of the experiment), provided their contributions could be discriminated from the Euler-Heisenberg signal. We briefly discuss how this could be achieved by exploiting microscopic details of the effective interaction, though a detailed analysis is left for future work. 

Firstly, the polarization structure of photon-photon scattering depends on symmetries of the mediating particle. For example, a pseudo-scalar field $a$ (e.g. describing axion-like particles) can only couple to the electromagnetic fields via the parity-odd invariant $\mathcal{G}_2$, i.e. $\mathcal{H}_{\rm int}\sim a \mathcal{G}_2$ resulting in an effective quartic interaction proportional to $(\mathcal{G}_2)^2$. In contrast, a scalar mediating field $\phi$ can only interact with photons via the parity-even invariant $\mathcal{G}_1$, i.e. $\mathcal{H}_{\rm int}\sim \phi \mathcal{G}_1$. As the two invariants depend differently on the photon polarization (e.g. $\mathcal{G}_2$ vanishes for linear polarizations), it follows that mapping out the full polarization structure of the forward-scattering signal is sufficient to discriminate these two cases from each other, as well as from the Euler-Heisenberg signal which has a distinct $4:7$ ratio of these two contributions. In general, the dependence on the polarization of the interferometric signal will depend on the structure of the microscopic interaction~\cite{Beckey2026}.

Secondly, the mass of the mediating particle $m$ dictates the locality of the effective photon-photon interaction. In the QED case, photon-photon interactions are effectively local due to the large separation between the single-photon energy (roughly $1$eV) and the electron mass ($\approx 0.5$MeV$/c^2$). For mediators with masses in the sub-eV regime, the effective interaction becomes non-local and sensitive to retardation effects. For ultra-light mediators, the interaction becomes non-local as the Compton wavelength $r_C=\hbar/(m c)$ of the mediator becomes comparable to the scale at which the optical intensity varies, e.g. $r_C$ is several centimeters for a mediator mass of order $\mu$eV$/c^2$. We previously studied photon-photon interactions mediated by massless gravitons in Ref.~\cite{Mehdi2023_QG} in the context of non-local interactions between counter-propagating beams separated by a distance $w$ in a ring-cavity geometry. By adopting such a geometry and varying $w$ one could distinguish a finite but small mediator mass $m$ where the interaction should vanish exponentially with $mc w/\hbar$ from the truly massless limit where the interaction scales as $\log(w)$~\cite{Mehdi2023_QG}.

For mediators with masses in the sub-eV regime, attenuation of the intra-cavity photon field could occur due to pair production in the case of charged fermions~\cite{Gies2006}, kinetic mixing with hidden-sector `dark photons'~\cite{Ahlers2008,Abel2008,Abel2008a}, or axion-photon conversion~\cite{Beyer2020}. In general, this attentuation will be polarization dependent leading to optical vacuum \emph{dichroism}. As vacuum dichroism predicted by QED is vanishingly small due to the large mass of the electron relative to the single-photon energy, a dichroic signal would indicate beyond-Standard-Model physics~\cite{Ejlli2020}. We leave a detailed theoretical analysis of dichroism in the context of the  experiments proposed here for future work.

\SectionPRL{Concluding remarks} In summary, this work presents an all-optical pathway to detect photon-photon interactions based on laser interferometry using standing-wave optical cavities. In particular, we have demonstrated that the detection of optical vacuum birefringence predicted by QED should be within reach of current state-of-the-art experiments. Our proposed experiment has several advantages over current approaches searching for vacuum nonlinearities: (1) it is compatible with conventional continuous-wave laser sources; (2) external magnetic fields are not required; (3) the achievable precision improves nonlinearly with intra-cavity power and can be made independent of cavity length; and (4) it leverages existing expertise and advanced technologies for high-power laser interferometry in the context of gravitational-wave detection. Indeed, efforts towards realizing the proposed experiment will shed invaluable light on technical noises and systematic effects of relevance to gravitational-wave detection, such as thermo-optical effects at the interface of the mirror surfaces and intra-cavity fields with circulating powers approaching the megawatt scale~\cite{Lu2024,Gras2015,Blair2017,Jia2021,Rosauer2025}.

In this work we have outlined several methods of discriminating photon-photon scattering from technical birefringence in the cavity, e.g. due to residual gas in the vacuum chamber via the Cotton-Mouton effect~\cite{Rizzo1997}. These include controlling the narrowband frequency response of the signal via intensity modulation, mapping out the full polarization dependence of the signal, and performing experiments with cavities of different lengths (exploiting the length-independence of the signal when the cavity volume is minimized). In principle, spurious signals due to mirror birefringence could be eliminated entirely by looking for interaction-induced cross-phase modulation between two independent cavities with overlapping intensity profiles, provided the mirrors of each cavity are suitably decoupled from one another (e.g. using independent suspension systems). Although the strength of the interaction would be smaller than that of a single-cavity scheme, for a given optical power, this could potentially be mitigated by optimizing the geometry such that the intensity in the overlapping region is maximized. We leave a detailed study of such an experiment for future work.

Beyond detection of photon-photon interactions predicted by QED, further improvements to the interferometric precision of the proposed experiment would enable increased sensitivity to certain beyond-Standard-Model quantum fields. In the near term, it would be most beneficial to increase the achievable circulating power in the cavity which could feasibly be several megawatts~\cite{Jia2021}. While phase squeezing could allow modest sensitivity gains, another appealing possibility is using multiple cavities that are simultaneously interrogated; for $N_c$ cavities with similar optical powers, one could could improve the sensitivity by $\sqrt{N_c}$ without entanglement or up to $N_c$ with entanglement using distributed sensing protocols~\cite{Beckey2026}. In the long term, robust detection schemes and improved detector efficiency could feasibly enable photon-counting schemes which can attain Heisenberg-scaling with bright optical sources, which would have unprecedented sensitivity that could enable all-optical tests of the quantization of gravity~\cite{Mehdi2023_QG}.

\emph{Acknowledgments.---} The authors acknowledge insightful discussions with Catalina Currceanu and Isabelle Savill-Brown, and are grateful to James Gardner and Giriraj Hiranandani for critical feedback on an early manuscript. We are additionally thankful to an anonymous referee of a previous manuscript~\cite{Mehdi2023_QG} for drawing our attention to Euler-Heisenberg photon-photon interactions. We acknowledge the Ngunnawal and Ngambri peoples as the original custodians of the land on which this research was conducted. Z.M. acknowledges support from the Australian Government through Australian Research Council Project No. DP260102306. S.A.H. acknowledges support through an Australian Research Council Future Fellowship, Grant No. FT210100495. 

\appendix

\section{\label{app:HamiltonianDerivation} Interaction Hamiltonian for forward scattering of photons in QED}

The quantized electric and magnetic fields in the non-interacting limit can be generated from the vector potential $\hat{A}_\mu = [0,\mathbf{\hat{A}}]$ in the Coulomb gauge: $\mathbf{\hat{E}}=-\partial_t \mathbf{\hat{A}}$ and $\mathbf{\hat{B}}=\nabla\times\mathbf{\hat{A}}$. Here $\mathbf{\hat{A}}$ can be expanded in terms of the annihilation ($\hat{a}_n$) and creation operators ($\hat{a}_n^\dag$) for the intra-cavity modes which satisfy the canonical commutation relation $[\hat{a}_n,\hat{a}_m^\dag]=\delta_{nm}$, i.e.
\begin{align}
\label{eq:app:QuantizedVectorPotential}
    \mathbf{\hat{A}}(\mathbf{r},t) = \sum_{n} \mathbf{e}_n\sqrt{\frac{\hbar}{\epsilon_0 \omega_n A_\perp}}\hat{a}_n u_n(z)e^{-i\omega_n t}+ {\rm h.c.} \,.
\end{align}
Here $\mathbf{e}_n$ is a normalized polarization vector transverse to the direction of propagation (taken to be the $z$-axis), $A_\perp$ is the effective transverse area of the cavity mode (related to the cavity volume by the cavity length $L$, i.e. $V=A_\perp L$), and $u_n(z)$ is the one-dimensional wavefunction of the intra-cavity field. Assuming a Fabry-Per\'ot cavity such that the intra-cavity modes are well-approximated as standing waves, $u_n(z)=\sqrt{2/L}\sin(k_n z)$ where $k_n=n\pi/L$ corresponds to a mode with (non-interacting) frequency $\omega_n = c k_n$. 

The derivation of $\hat{H}_{\rm SM}$ follows by substitution of the quantized EM field for a single optical mode with annihilation operator by $\hat{a}$ and polarization $\mathbf{e}=\{e_x,e_y,0\}$ into Eq.~\eqref{eq:HamiltonianDensity}, normally ordering the operator strings, making an optical rotating wave approximation, and finally integrating over the cavity volume. Details of these steps are provided in the Supplemental Materials~\cite{SupplementalMaterials}. The result can be expressed in terms of the contributions of the two Lorentz invariants:
\begin{align}
    \int d^3x \; \hat{\mathcal{G}}^2_1     &= \frac{2(\hbar \omega_0)^2\hat{a}^\dag\hat{a}^\dag\hat{a}\hat{a}}{V \epsilon_0^2}\left(1+|\mathbf{e}\cdot\mathbf{e}|^2 \right)\\ 
    \int d^3x \; \hat{\mathcal{G}}^2_2     &= 8(\Im{e_x e_y^*})^2\frac{(\hbar \omega_0)^2\hat{a}^\dag\hat{a}^\dag\hat{a}\hat{a}}{V\epsilon_0^2} \,.
\end{align}
The polarization dependence in the above expressions can be recast in terms of the normalization condition $|\mathbf{e}^*\cdot\mathbf{e}|^2=1$ and the circular Stokes parameter $\mathcal{V} = 2\Im{e_x^*e_y}$, e.g. $|\mathbf{e}\cdot\mathbf{e}|^2 =1-\mathcal{V}^2$. Adding up the above contributions according to Eq.~\eqref{eq:HamiltonianDensity}, i.e.
\begin{align}
    \hat{H}_{\rm SM} = - \frac{2\alpha^2\varepsilon_0^2\hbar^3}{45m_e^4c^5}\int d^3x\left(4 \hat{\mathcal{G}}^2_1 + 7\hat{\mathcal{G}}^2_2\right) \,,
\end{align}
gives Equation \eqref{eq:Hint_SM} of the main text. The derivation of the two-mode expression [$\hat{H}_{\rm TM}=-\hbar\lambda\hat{n}_1\hat{n}_2$] follows the same steps but is more algebraically involved. Details of this calculation are provided in the Supplemental Materials for completeness~\cite{SupplementalMaterials}.

\section{\label{app:CavityVolume} Minimum cavity volume}
The volume of an optical mode in a Fabry-Perot cavity can be estimated by integrating the cross-sectional area of the beam over the cavity length. For simplicity, we model the transverse profile of the beam as circular, with a radius given by the $1/{\rm e}^2$ width of a Gaussian beam (which has the least divergence of any optical mode): 
\begin{align}
    w(z)^2 = w_0^2\left(1+\frac{z^2}{z_0^2} \right)
\end{align}
where $z_0 = \pi w_0^2/\lambda$ is the Rayleigh length of the beam for a laser wavelength $\lambda$ for a cavity aligned along the $z$ axis centered at $z=0$. Integrating $\pi w(z)^2$ over the cavity length $L$ yields the cavity volume as a function of beam waist and length:
\begin{align}
    V = L\pi w_0^2 + \frac{L^3 \lambda^2}{12\pi w_0^2} \,.
\end{align}
The first term in this expression gives the cavity volume for a perfectly collimated beam with radius $w_0$, and the second term accounts for optical diffraction over the cavity length $L$. The smallest possible cavity volume for a fixed $L$ can be obtained by minimizing $V$ with respect to $w_0$. The value $w_0^2={\lambda L/(2\pi\sqrt{3})}$ gives the minimum cavity volume:
\begin{align}
    V \geq V_{\rm min} = \frac{\lambda L^2}{\sqrt{3}} \,,
\end{align}
which is the result quoted in the main text. 

\section{Derivation of noise PSD for estimating the per-particle interaction rate \label{app:NoisePSD}}

Here we derive the PSD associated with estimating $\lambda(\Theta)$ from phase-sensitive measurements of mode $1$ (the `probe' mode) induced by macroscopic occupation of mode $2$ (the `pump' mode). We work in mean-field limit where we can make the replacement $\hat{n}_2\rightarrow \bar{n}_{\rm pump}(t)$, which we assume to be time-dependent, e.g. in the case of intensity modulation. Linearizing the optical fluctuations around a (real-valued) coherent state amplitude of the probe mode, i.e. $\hat{a}_1\rightarrow \hat{a} +\sqrt{\bar{n}_{\rm probe}}$ where $\hat{a}$ is annihilation operator for the vacuum of the pump mode, the two-mode interaction Hamiltonian can be approximated as:
\begin{align}
   \hat{H}_{\rm eff}=-\hbar\mathcal{J}(t)\hat{X} \,,
\end{align}
where $\mathcal{J}(t) \equiv \bar{n}_{\rm pump}(t)\sqrt{2\bar{n}_{\rm probe}} \lambda$ and $\hat{X}\equiv (\hat{a}+\hat{a}^\dag)/\sqrt{2}$. The complementary observable is the phase quadrature operator $\hat{Y}\equiv -i(\hat{a}^\dag-\hat{a})/\sqrt{2}$, which satisfies the canonical commutation relation $[\hat{X},\hat{Y}]=i$. For the effective Hamiltonian above, the equations of motion for the cavity field operators in the co-rotating frame with respect to the cavity frequency are~\cite{Beckey2026}:
\begin{subequations}
\label{eq:HeisenbergLangevinEqns}
    \begin{align}
    \frac{d\hat{X}}{dt} &= -\frac{\kappa}{2}\hat{X}+\sqrt{\kappa}\hat{X}^{\rm in} \\
    \frac{d\hat{Y}}{dt} &=  \mathcal{J}(t)-\frac{\kappa}{2}\hat{Y}+\sqrt{\kappa}\hat{Y}^{\rm in}
\end{align}
\end{subequations}
where $\hat{X}^{\rm in},\hat{Y}^{\rm in}$ are quadratures of the input field transmitted through the partially-reflective mirror at rate $\kappa$ (assuming an overcoupled cavity for simplicity) with zero mean. The phase of the output field transmitted from the cavity is related to the cavity field and input field by the boundary condition $\hat{Y}^{\rm out} = \hat{Y}^{\rm in}-\sqrt{\kappa}\hat{Y}$~\cite{Beckey2026}. We can obtain this field analytically by solving Eq.~\eqref{eq:HeisenbergLangevinEqns} in the frequency domain  [$\hat{Y}\rightarrow \hat{Y}(\omega)$], i.e.
\begin{align}
  \hat{Y}_m^{\rm out}(\omega) =& -\sqrt{\kappa}\chi_c(\omega)\mathcal{J}(\omega) + \left(1-\kappa\chi_c(\omega)\right)\hat{Y}_m^{\rm in}(\omega)  \label{eq:YOUT_FreqSpace}
\end{align}
in terms of the cavity susceptibility $\chi_c(\omega) \equiv [-i\omega+\kappa/2]^{-1}$. The corresponding noise PSD can be obtained using the Wiener-Khinchin theorem 
\begin{align}
2\pi S_{YY}(\omega)\delta(\omega-\omega')=\langle \delta\hat{Y}^{\rm out}(\omega)\delta\hat{Y}^{\rm out}(\omega')\rangle    
\end{align}
where $\delta \hat{Y}^{\rm out}\equiv \hat{Y}^{\rm out}|_{\mathcal{J}\rightarrow 0}$ denotes the phase field in the absence of the signal $\mathcal{J}(t)$. For laser light without squeezing $S_{YY}^{\rm in}(\omega)=1/2$, which gives the output noise PSD $S_{YY}(\omega)=1/2$. If the phase quadrature of the input beam is squeezed, then $S_{YY}^{\rm in}(\omega)=e^{-2r}/2$ where $r>0$ is a real-valued parameter. The noise PSD for the estimate of the interaction parameter $\lambda$ can then be obtained rescaling $S_{YY} \rightarrow S_{YY}/(2\kappa\bar{n}_{\rm probe} \bar{n}_{\rm pump}^2| \chi_c|^2)$, which is implicitly evaluated at the modulation frequency of the pump mode. Then, relabelling the `probe' as mode $i$ and the `pump' as mode $j$, the PSD Equation \eqref{eq:PSD_ij} of the main text.

\section{Quantum Fisher information for a two-mode squeezed vacuum state\label{app:QFI}}
Here we calculate the quantum Fisher information (QFI) associated with estimating the strength of an interaction of the form $\hat{H}=-\hbar\lambda\hat{n}_1\hat{n}_2$. Over the cavity lifetime $\tau$, the parameter $\lambda$ is encoded on an initial quantum state $\ket{\Psi}$ by the generator $\hat{G}= \hat{n}_1\hat{n}_2 \tau$. Assuming the state remains pure, the QFI can then be expressed as $\mathcal{I}_Q=4{\rm Var}(\hat{G}) = 4{\rm Var}(\hat{n}_1\hat{n}_2)\tau^2$. The form of the generator implies high sensitivity to states with significant phase correlations between the two modes, which motivates us to consider the two-mode squeezed vacuum state
\begin{align}
 \ket{{\rm TMSV}} =\hat{U}(re^{i\theta})\ket{0} \,,
\end{align}
where $\hat{U}(\chi)=\exp{\frac{1}{2}\left(\chi^*\hat{a}_1\hat{a}_2-\chi\hat{a}_1^\dag\hat{a}_2^\dag\right)}$ describes two-mode squeezing with angle $\theta$ and magnitude $r$ for $\chi = re^{i\theta}$. We take $\theta=0$ for simplicity. We work in the Heisenberg picture where the cavity operators are transformed as:
\begin{align}
    \hat{a}_1 &\rightarrow \hat{a}_1 \cosh(r) + \hat{a}_2^\dag \sinh(r) \,,\\
   \hat{a}_2 &\rightarrow \hat{a}_2 \cosh(r) + \hat{a}_1^\dag \sinh(r) \,.
\end{align}
The resulting expectation values are taken with respect to the two-mode vacuum. To simplify this calculation we normally order the operator strings using $[\hat{a},\hat{a}^\dag]=1$, such that only non-operator-valued contributions contribute to the expectation values. This is used to compute the variance of the generator:
\begin{align}
   \notag {\rm Var}(\hat{n}_1\hat{n}_2)&=\frac{1}{4} \sinh ^2(r) (\cosh (2 r)+3 \cosh (6 r)) \,, \\
    &= N+13N^2 + 32 N^3 + 20N^4 \,,
\end{align}
in terms of the mean photon number in the cavity, $N=\langle \hat{n}_1+\hat{n}_2\rangle = \sinh^2(r)$. This leads to a QFI of approximately $80N^4\tau^2$. This can be related to the quantum Cramer-Rao bound $\Delta \lambda^2 \geq 1/(M\mathcal{I}_Q)$ for $M\approx T/\tau$ cavity lifetimes during a total integration time $T$, which gives the noise limit $\Delta \lambda \geq [80N^4( \tau T)]^{-1/2}$.

\bibliography{Bibliography}

\include{SupplementalMaterials}

\end{document}

%% file: SupplementalMaterials.tex


\onecolumngrid
\clearpage
\setcounter{secnumdepth}{2}

\setcounter{equation}{0}\renewcommand{\theequation}{S\arabic{equation}}
\setcounter{figure}{0}  \renewcommand{\thefigure}{S\arabic{figure}}
\setcounter{table}{0}   \renewcommand{\thetable}{S\arabic{table}}
\setcounter{section}{0} \renewcommand{\thesection}{S\Roman{section}}
\setcounter{page}{1}\renewcommand{\thepage}{S\arabic{page}}

\setcounter{equation}{0}\renewcommand{\theequation}{S\arabic{equation}}
\setcounter{section}{0} \renewcommand{\thesection}{S\Roman{section}}

\begin{center}
  {\large\bfseries Supplemental Materials: Quantum Vacuum Nonlinearities in Laser Interferometers \par }
  \vspace{1em}
  Zain Mehdi$^{1}$, Joseph J. Hope$^{1}$, Simon A. Haine$^{1}$\par
  \vspace{0.5em}
  {\itshape $^{1}$Department of Quantum Science and Technology and Department of Fundamental and Theoretical Physics, Research School of Physics, Australian National University, Canberra 2600, Australia\par}
  \vspace{0.5em}
  (Dated: \today)
\end{center}
\vspace{1em}

In this supplemental document we provide additional details of (I) how the leading-order interaction Hamiltonian is obtained from the low-energy limit of Euler-Heisenberg effective field theory, and (II) the derivation of the few-mode interaction Hamiltonians, $\hat{H}_{\rm SM}$ and $\hat{H}_{\rm TM}$, studied in the main text.

\section*{I: Leading-order interaction Hamiltonian from the Euler-Heisenberg Lagrangian}
The starting point of this analysis is the low-energy limit of the Euler-Heisenberg Lagrangian, which includes photon-photon interactions at single-loop order in QED (repeated indices are implicitly summed):
\begin{align}
    \mathcal{L} = \underbrace{\frac{1}{2}(E^2-B^2)}_{\equiv \mathcal{L}^{(0)}}+\epsilon\left( \underbrace{(E^2-B^2)^2+7(E_iB_i)^2}_{\equiv\mathcal{L}^{(1)}}\right) = \mathcal{L}^{(0)}+ \epsilon \mathcal{L}^{(1)}
\end{align}
where $\epsilon = 2\alpha^2/(45 m_e^4) \ll 1$ is our perturbative parameter, $E^2\equiv E_iE_i$ and $B^2=B_iB_i$. We have chosen to work in natural units ($\hbar=c=1$). 
Here we have separated the free-field EM Lagrangian density $\mathcal{L}^{(0)}$ from the interaction term $\mathcal{L}^{(1)}$. The corresponding Hamiltonian density can be obtained by performing a Legendre transformation on $\mathcal{L}$, i.e.
\begin{align}
\label{eq:Supp:LEGENDRETRANSFORM}
    \mathcal{H} =  \Pi^i(\partial_t A_i) - \mathcal{L}^{(0)}- \epsilon \mathcal{L}^{(1)} =  -\Pi^iE_i - \mathcal{L}^{(0)}- \epsilon \mathcal{L}^{(1)}\,,
\end{align}
where $\partial_t A_i = -E_i$ ($i=1,2,3$) [$A_0=0$ in the matter vacuum] and $\Pi_i$ are the canonical field momenta:
\begin{align}
    \Pi^i \equiv \frac{\partial \mathcal{L}}{\partial(\partial_t A_i)} = -\frac{\partial \mathcal{L}}{\partial E_i} = -\left( E_i + \epsilon f_i(\vec{E},\vec{B})\right) \,,
\end{align}
where $\vec{E}=(E_x,E_y,E_z)$ and similarly for $\vec{B}$. Note that we do not distinguish between raised and lowered indices on the spatial components, corresponding to choosing the metric signature $[-,+,+,+ ]$. As compared to the free-field case (where $\Pi^i=-E^i$), the canonical momenta gets a correction proportional to the EM polarization tensor:
\begin{align}
   f_i(\vec{E},\vec{B}) = \frac{\partial\mathcal{L}^{(1)}}{\partial E_i} = 4E_i(E^2-B^2)^2 + 14 B_i(E_jB_j)\,.
\end{align}
To clearly identify the interaction term in the Hamiltonian density, we write $\mathcal{H}$ in terms of the canonical variables up to $\mathcal{O}(\epsilon)$ using the inversion $E^i = -\Pi^i - \epsilon f^i(\vec{E},\vec{B}) = -\Pi^i - \epsilon f^i(\vec{\Pi},\vec{B})+\mathcal{O}(\epsilon ^2)$:
\begin{align}
    \mathcal{H} &= \vec{\Pi}\cdot\left(\vec{\Pi}+\epsilon \vec{f}\right) - \frac{1}{2}\left( [\vec{\Pi}+\epsilon\vec{f} ]\cdot [\vec{\Pi}+\epsilon\vec{f} ] - B^2\right) - \epsilon\left( (\Pi^2-B^2)^2+7(\vec{\Pi}\cdot\vec{B})^2\right)+ \mathcal{O}(\epsilon^2) \, \\
    &= \underbrace{\frac{1}{2}\left(\Pi^2+B^2\right)}_{\equiv \mathcal{H}_0} -\epsilon\left( (\Pi^2-B^2)^2+7(\vec{\Pi}\cdot\vec{B})^2\right)+ \mathcal{O}(\epsilon^2)\,.
\end{align}
The first term describes linear evolution of the system (after canonical quantization, $\hat{H}_0 = \int d^3x \mathcal{H}_0 = \sum_k \hbar \omega_k \hat{a}_k^\dag \hat{a}_k$ up to a constant~\cite{Hillery1984}). The remaining term describes the nonlinear field dynamics, leading to the following interaction Hamiltonian density to leading order in $\epsilon$:
\begin{align}
    \mathcal{H}_{\rm int} = -\epsilon\left( (E^2-B^2)^2+7(\vec{E}\cdot\vec{B})^2\right)+ \mathcal{O}(\epsilon^2) \,.
\end{align}
Note that this is precisely the negative of the interaction term in the Euler-Heisenberg Lagrangian density, which we could have (wrongly) obtained by ignoring the correction to the free-field canonical momenta (i.e. taking $\Pi_i = -E_i$) and taking $\mathcal{H}_{\rm int}=-\epsilon\mathcal{L}^{(1)}$. This means that, to leading order in perturbation theory, we can simply substitute the quantized free-field EM field into the Euler-Heisenberg interaction to obtain the quantized effective theory to $\mathcal{O}(\epsilon)$. This is the approach taken in Appendix A of the main text. Higher-order corrections can be obtained by following the formal procedure for quantizing the EM field in a nonlinear medium~\cite{Hillery2009}.

\section*{II: Detailed derivation of the few-mode interaction Hamiltonians}
As described in Appendix A of the main text, we obtain the effective photon-photon interaction Hamiltonian for the intra-cavity field by substituting the quantized free EM fields into $\hat{H}_{\rm int} = \int d^3x\;\mathcal{H}_{\rm int}$ following a normal-ordering prescription, including only modes that satisfy the boundary conditions of the cavity mirrors. We will focus on the resonant terms involving only modes with macroscopic occupation which generate the interferometric signals that we study in the main text.

\textbf{A single optical mode.---} For the case of a single optical mode with annihilation operator $\hat{a}$ (satisfying the bosonic canonical commutation relation $[\hat{a},\hat{a}^\dag]=1$) and frequency $\omega$, the quantized EM fields can be written in terms of its polarization vector $\mathbf{e}=(e_x,e_y,0)$:
\begin{align}
\label{eq:SM_EandBfields}
    \hat{\mathbf{E}} &= i\sqrt{\frac{2}{\omega V}} \sin( \omega z)( \hat{a}^\dag e^{i\omega t} \mathbf{e}^* -\hat{a} e^{-i\omega t} \mathbf{e})\\
     \hat{\mathbf{B}} &=  \sqrt{\frac{2\omega}{V}} \cos( \omega z)( \hat{a} e^{-i\omega t} \mathbf{e}_z\times\mathbf{e} + \hat{a}^\dag e^{i\omega t}(\mathbf{e}_z\times\mathbf{e})^* )
\end{align}
where $\mathbf{e}_z=(0,0,1)$ is the unit vector along the cavity axis ($\mathbf{e}\cdot \mathbf{e}_z=0$ by construction), and $V$ is the cavity volume. The factor of $\sqrt{2}$ comes from the normalization of the standing wave. The time dependence of these operators follows from working in the interaction picture with respect to $\hat{H}_0=\hbar\omega(\hat{a}^\dag\hat{a}+1/2)$.

Rather than substituting Equation \eqref{eq:SM_EandBfields} directly into $\mathcal{H}_{\rm int}$, to highlight the polarization structure of the interaction we separately evaluate contributions of the two Lorentz invariants $\mathcal{G}_1\equiv (E^2-B^2)/2$ and $\mathcal{G}_2 = E_iB_i$ to the interaction Hamiltonian. In each case we expect to get terms proportional to $\hat{a}^\dag\hat{a}^\dag\hat{a}\hat{a}$, as any other contribution rotate at multiples of $\omega$ and thus get discarded in a rotating wave approximation. The pre-factors to these terms will involve integrals over the cavity modes, where we will approximate the transverse integral as a flat-top function with total area $A_\perp=V/L$. We will also make use of the following results for the overlap of standing waves:
\begin{subequations}
\label{eq:Supp:SM_Integrals}
    \begin{align}
   & \int_0^L dz \sin^4(\omega z) = \int_0^L dz \cos^4(\omega z) = \frac{3L}{8} \,, \\
    &     \int_0^L dz \sin^2(\omega z) \cos^2(\omega z)  = \frac{L}{8} \,,\\
    &    \int_0^L dz \sin^2(\omega z) =  \int_0^L dz \sin^2(\omega z) = \frac{L}{2} \,,
\end{align}
\end{subequations}

where it is implicitly assumed that $\omega = n\pi/L$ ($n$ is an integer).

Starting with the first Lorentz invariant, we can use Eq.~\eqref{eq:SM_EandBfields} and the integrals above to obtain the following expression within an optical rotating-wave approximation (where $:\star:$ denotes normal ordering of $\star$):
\begin{align}
  \notag  \int d^3x : \hat{\mathcal{G}}_1^2 : &= \frac{\omega^2\hat{a}^\dag\hat{a}^\dag\hat{a}\hat{a}}{V \epsilon_0^2}\left(2 e_x^2 \left(\left(e_y\right){}^*\right){}^2+4 e_x e_y \left(e_x\right){}^* \left(e_y\right){}^*+2 e_y^2 \left(\left(e_x\right){}^*\right){}^2+4 e_x^2 \left(\left(e_x\right){}^*\right){}^2+4 e_y^2 \left(\left(e_y\right){}^*\right){}^2\right) \,,\\
    &= \frac{\omega^2\hat{a}^\dag\hat{a}^\dag\hat{a}\hat{a}}{V \epsilon_0^2}\left( 2(|e_x|^2+|e_y|^2)^2 + 2|e_x^2+e_y^2|^2 \right)
    = \frac{2\omega^2\hat{a}^\dag\hat{a}^\dag\hat{a}\hat{a}}{V \epsilon_0^2}\left(1+|\mathbf{e}\cdot\mathbf{e}|^2 \right) \,.
\end{align}
In the second line we have used the normalization of the polarization vector, $|e_x|^2+|e_y|^2=1$. We can re-write this in terms of the circular Stokes parameter, $\mathcal{V} \equiv -2\Im{e_x e_y^*}$, which it tells you how much net helicity there is in the beam (for linear polarizations $\mathcal{V}=0$, for circular $\mathcal{V}=\pm 1$), which gives
\begin{align}
\label{eq:Supp:SM_FirstInvariantSquared}
   \int d^3x : \hat{\mathcal{G}}_1^2: &= (4-2\mathcal{V}^2)\frac{\omega^2\hat{a}^\dag\hat{a}^\dag\hat{a}\hat{a}}{V}\,.
\end{align}
Applying the same procedure to the second invariant term yields 
\begin{align}
    \int dV :\hat{\mathcal{G}}_2^2 : &= -2\frac{\omega^2\hat{a}^\dag\hat{a}^\dag\hat{a}\hat{a}}{V} \left(e_x \left(e_y\right){}^*-e_y \left(e_x\right){}^*\right){}^2 \\
    &= 8\Im{e_x e_y^*}^2\frac{\omega^2\hat{a}^\dag\hat{a}^\dag\hat{a}\hat{a}}{V} = 2\mathcal{V}^2\frac{\omega^2\hat{a}^\dag\hat{a}^\dag\hat{a}\hat{a}}{V} \,.\label{eq:Supp:SM_Invariant2Squared}
\end{align}
Notably, this contribution is only non-zero if the polarization is not (exactly) linear, i.e. if $\mathcal{V}\neq 0$.

Combining Equations \eqref{eq:Supp:SM_FirstInvariantSquared} and \eqref{eq:Supp:SM_Invariant2Squared}, we arrive at the result for the single-mode interaction Hamiltonian:
\begin{align}
    \hat{H}_{\rm SM} &= -\epsilon\int d^3x\left(4:[\mathcal{G}_1]^2:+7:[\mathcal{G}_2]^2:\right) =-\frac{2\alpha^2}{45m_e^4}\frac{\omega^2\hat{a}^\dag\hat{a}^\dag\hat{a}\hat{a}}{V}  \left(4[4-2\mathcal{V}^2]+7[2\mathcal{V}^2] \right)\,, \\
    & = - (1+3\mathcal{V}^2/8)\frac{32\alpha^2}{45m_e^4}\frac{\omega^2\hat{a}^\dag\hat{a}^\dag\hat{a}\hat{a}}{V} \,.
\end{align}
Restoring SI units by multiplying this expression by a factor of $(\hbar/c)^5$, this gives $\hat{H}_{\rm SM}$ in the main text.

\textbf{Two-mode interactions.---} Next, we consider interactions between two macroscopically occupied modes $\hat{a}_1$ and $\hat{a}_2$ of the cavity, such that the quantized EM fields can be expressed by generalizing Eq.~\eqref{eq:SM_EandBfields}:
\begin{align}
\label{eq:TM_EandBfields}
    \hat{\mathbf{E}} &= i\sum_{n=1,2} \sqrt{\frac{2}{\omega_n V}}\sin( \omega_n z)( \hat{a}_n^\dag e^{i\omega_n t} \mathbf{e}_n^* -\hat{a}_n e^{-i\omega_n t} \mathbf{e}_n)\\
     \hat{\mathbf{B}} &=  \sum_{n=1,2}\sqrt{\frac{2\omega_n}{V}} \cos( \omega_n z)( \hat{a}_n e^{-i\omega_n t} \mathbf{e}_z\times\mathbf{e}_n + \hat{a}_n^\dag e^{i\omega_n t}(\mathbf{e}_z\times\mathbf{e}_n)^* ) \,.
\end{align}
When substituting these fields into $\hat{H}_{\rm int}$, integrals arise that depend on whether or not the two modes modes are degenerate (e.g. two orthogonal polarizations of the same frequency) or two separate resonances of the cavity. The following integrals arise:
\begin{subequations}
    \begin{align}
    &\int_0^L  \sin(\omega_1 z)\sin(\omega_2 z) \cos(\omega_1 z)\cos(\omega_2 z)dz = \frac{L}{8}\delta_{n_1,n_2}  \,, \\
    &\int_0^L \sin^2 (\omega_1z)\,\sin^2(\omega_2 z)\,dz
=\int_0^L \cos^2(\omega_1 z)\,\cos^2(\omega_2 z)\,dz
=\frac{L}{4}+\frac{L}{8}\,\delta_{n_1,n_2} \,,\\
&\int_0^L \cos^2(\omega_1 z)\,\sin^2(\omega_2 z)\,dz
=\frac{L}{4}-\frac{L}{8}\,\delta_{n_1,n_2} \,,
\end{align}
\end{subequations}
where $\omega_1 = n_1\pi/L$ and $\omega_2 = n_2\pi/L$. It is straightforward to confirm that the single-mode integrals (Eq.~\eqref{eq:Supp:SM_Integrals}) are recovered in the limiting case $n_1=n_2$. Using these results, we find the following contributions of the two Lorentz invariants $(\hat{\mathcal{G}}_{1,2})^2$ to interaction Hamiltonian (ignoring the single-mode terms proportional to $\hat{a}_i^\dag\hat{a}^\dag _i\hat{a}_i\hat{a}_i$ for $i=1,2$):
\begin{subequations}
\label{eq:TM_Invariants_Contributions}
    \begin{align}
\label{eq:TM_Invariant_FF}
    \int d^3x :\hat{\mathcal{G}}_1^2: &= \frac{\omega_1\omega_2 \hat{n}_1\hat{n}_2}{V}\big[  4 \left(\left| \mathbf{e}_1^*\cdot\mathbf{e}_2\right| ^2+| \mathbf{e}_1\cdot\mathbf{e}_2| ^2\right)  +4\delta_{12} \left(1+|\mathbf{e}_1\cdot\mathbf{e}_2| ^2 \right)    \big] \,, \\
    \label{eq:TM_Invariant_GG}
    \int d^3x :\hat{\mathcal{G}}_2^2: &= \frac{\omega_1\omega_2 \hat{n}_1\hat{n}_2}{V}\big[  4 \left(2-\left| \mathbf{e}_1^*\cdot\mathbf{e}_2\right| ^2-| \mathbf{e}_1\cdot\mathbf{e}_2| ^2 \right)  +4\delta_{12} \left(2\left| \mathbf{e}_1^*\cdot\mathbf{e}_2\right| ^2-| \mathbf{e}_1\cdot\mathbf{e}_2| ^2-1 \right)    \big] \,.
\end{align} 
\end{subequations}
where $\hat{n}_i=\hat{a}^\dag_i\hat{a}_i$ is the photon number operator ($i=1,2$). Using Equation \eqref{eq:TM_Invariants_Contributions}, we can obtain the two-mode interaction Hamiltonian used in the main text:
\begin{align}
    \hat{H}_{\rm TM} &= -\epsilon\int d^3x\left(4:[\mathcal{G}_1]^2:+7:[\mathcal{G}_2]^2:\right) \,, \\
    &=  -\frac{\epsilon\omega_1\omega_2 \hat{n}_1\hat{n}_2}{V} \left(4\big[  4 \left(a+b\right)  +4\delta_{12} \left(1+b \right)    \big] + 7 \big[  4 \left(2-a-b \right)  +4\delta_{12} \left(2a-b-1 \right)    \big] \right) \\
    &=  -\frac{8\alpha^2\omega_1\omega_2 \hat{n}_1\hat{n}_2}{45V m_e^4} \left([14-3a-3b]+\delta_{n_1,n_2}\left[14a-3(b+1)\right] \right)
\end{align}
where we have adopted the shorthands $a=\left| \mathbf{e}_1^*\cdot\mathbf{e}_2\right| ^2$ and $b=| \mathbf{e}_1\cdot\mathbf{e}_2| ^2$, and substituted $\epsilon = 2\alpha^2/(45m_e^4)$ in the last line. Conversion to SI units follows by multiplying the pre-factor by $(\hbar/c)^5$ as in the single-mode case, which gives the form of $\hat{H}_{\rm TM}$ quoted in the main text.

\emph{Remarks.---} To obtain the form of $\hat{H}_{\rm TM}$ above, we have performed an optical rotating-wave approximation to discard four-wave mixing terms of the form $\hat{a}_1^\dag \hat{a}_2\hat{a}_2\hat{a}_2 e^{i(\omega_1-3\omega_2)t}+{\rm h.c.}$ under the assumption that $\omega_2\neq 3\omega_1$. We have also neglected mode-mixing terms of the form $\hat{a}_1^\dag \hat{a}_1\hat{a}_2^\dag\hat{a}_1 e^{i(\omega_2-\omega_1)t} + {\rm h.c.}$, as these do not contribute significantly to the interferometric readouts we consider in the main text which are sensitive to phase shifts generated by $\hat{n}_i^2$ and $\hat{n}_1\hat{n}_2$. In the non-degenerate case, discarding these terms can be justified by a rotating-wave approximation given interrogation times that are long compared to $|\omega_1-\omega_2|^{-1}$. While this approximation cannot be made in the degenerate case, the approximation could alternatively be justified by noting mode-mixing generators of the form $\epsilon \hat{a}_1^\dag \hat{a}_2$ only generate dynamics at $\mathcal{O}(\epsilon^2)$ assuming the input beam to be in a two-mode coherent state. Such terms could be significant in photon-number-counting readouts with non-classical states of light~\cite{Mehdi2023_QG}, which we will explore in a future work.





